\documentclass[%
superscriptaddress,
preprint,
 amsmath,amssymb,
 aps,
]{revtex4-2}

\usepackage{graphicx}
\usepackage{dcolumn}
\usepackage{bm}
\usepackage{color}

\begin{document}

\title{Augmented Magnetic Octupole in Kagom\'{e} Antiferromagnets Detectable via X-ray Magnetic Circular Dichroism}

\author{Yuichi Yamasaki}
\affiliation{Research and Services Division of Materials Data and Integrated System (MaDIS), National Institute for Materials Science (NIMS), Tsukuba, 305-0047, Japan}
\affiliation{RIKEN Center for Emergent Matter Science (CEMS), Wako 351-0198, Japan
}
\affiliation{PRESTO, Japan Science and Technology Agency (JST)}
\author{Hironori Nakao}
\affiliation{
Condensed Matter Research Center and Photon Factory, Institute of Materials Structure Science, High Energy Accelerator Research Organization, Tsukuba 305-0801, Japan
}
\author{Taka-hisa Arima}
\affiliation{RIKEN Center for Emergent Matter Science (CEMS), Wako 351-0198, Japan
}
\affiliation{
Department of Advanced Materials Science, University of Tokyo, Kashiwa 277-8561, Japan
}
\maketitle

\textbf{
The magneto-optical Kerr effect (MOKE) has recently been discovered in antiferromagnetic Kagom\'{e} lattice Mn$_3$Sn \cite{Mn3Sn_Kerr_higo2018large}. 
Since the compound exhibits a coplanar $120^\circ$ antiferromagnetic (AFM) order, the magnetic moments cancel each other, and the net magnetization is almost zero. 
However, the MOKE is allowed due to the lack of the time reversal symmetry (TRS) of the AFM order.
X-ray magnetic circular dichroism (XMCD) can detect a difference in up and down spin density of states (DOS), and thus XMCD originating from the spin operator, the so-called $S_z$ term, should be negligibly small as well as the net magnetization. 
Nonetheless, we show that XMCD originating from the magnetic dipole operator, the so-called $T_z$ term, should remain uncancelled in the AFM with the broken TRS. 
Moreover, we clarify that there is a strong link between the $T_z$ term and the augmented magnetic octupole (AMO), which suggests that XMCD can be an effective approach for analyzing multipole moments in antiferromagnets.
}

Hexagonal Mn$_3$Sn with magnetic Mn ions  on the breathing Kagom\'{e} lattice shows the coplanar 120$^\circ$ AFM order which gives rise to the large thermoelectric effects \cite{ikhlas2017large, Mn3Sn_Nernst}, anomalous Hall effect \cite{Mn3Sn_Nakatsuji}, and inverse spin Hall effect \cite{Mn3Sn_ISHE}.
It is identified as a magnetic Weyl semimetal, where the Dirac points at band crossing are separated into two sets of Weyl points owing to the broken TRS.
The Weyl points, examined by theoretical calculations \cite{Mn3Ir_Mac, kuroda2017evidence} and by transport measurements \cite{Mn3Sn_Nakatsuji}, exhibit nontrivial transport anomalies because of the non-zero Berry curvature.
A slight distortion from the perfect 120$^\circ$ AFM order leaves small magnitude of residual ferromagnetism, however the huge MOKE seems to respond to the AFM structure \cite{Mn3Sn_Kerr_higo2018large}.
The concept of AMO, a magnetic cluster composed of a Mn triangle of the same symmetry as the magnetic octupole, has been introduced to elucidate those phenomena \cite{Mn3Sn_MTSuzuki}.
Although the AMOs may express XMCD because of the broken TRS, it is not yet clear whether AMO considered in a multiatom cluster can be detected by XMCD observed via the absorption process of each atom.
Although the detection of the magnetic octupole moment on an atom via the quadrupole-quadrupole ($E2$-$E2$) process of resonant x-ray scattering has been discussed \cite{joly2012resonant,matsumura2009magnetic}, here we show that the AMO in Mn$_3$Sn is observable as the $T_z$ term via the dipole-dipole ($E1$-$E1$) process of XMCD.

XMCD is an experimental technique that uses circularly polarized x-rays to obtain the magnetic information \cite{XMCD_first,XMCD_Shutz,lovesey1996x}.
It is maximized when the magnetization $\bm{m}$ is parallel to the photon wave vector $\bm{\hat{k}}$, where the electric field of light $\bm{E}$ perpendicular to $\bm{m}$.
However, it has different values depending on the orientation of electron orbital even with the same $\bm{m}$ \cite{STOHR1995253}; for example, XMCD signals differ between $3z^2-r^2$ and $3y^2-r^2$ orbitals for $3d$ electron even under the same condition of $\bm{m} || \bm{k} || z$ [Figs. 1\textbf{a} and \textbf{b}].
It comes from a difference in photon absorption coefficients between $E||x$ and $E||y$; $i.e.$ isotropic for the $3z^2-r^2$ orbital but anisotropic for the $3y^2-r^2$ orbital.
The orbital orientation dependent XMCD is related to the spin density distribution, as discussed in more detail below. 

The difference is also manifested experimentally as the so-called $T_z$ term in the sum rule of XMCD, which has been well known as a powerful tool for separately analyzing the angular momenta of spin and orbital \cite{PhysRevLett.68.1943,PhysRevLett.70.694}.
If the $z$ axis is set to be parallel with $\bm{\hat{k}}$, using XMCD signal at $L_3$-edge ($\Delta_{L_3}$) and $L_2$-edge ($\Delta_{L_2}$), the total sum can be written as
\begin{align}
    \frac{\Delta_{L_3}+\Delta_{L_2}}{\Sigma} &= -\frac{\langle L_z \rangle}{2},
\end{align}
with $\Sigma$ and $\langle L_z \rangle$ being the energy integration of the x-ray absorption spectrum and the expectation value of the orbital operator $L_z$ for the $3d$ state, respectively.
Equation (1) indicates that the XMCD sum is only proportional to $\langle L_z \rangle$, and thus the orbital angular moment can be directly determined.
The other expression of the sum rule is written as, 

\begin{align}\label{sum_rule_2}
    \frac{\Delta_{L_3}-2\Delta_{L_2}}{\Sigma} &= \frac{2}{3}\langle S_z \rangle +\frac{7}{3} \langle T_z \rangle, 
\end{align}
with $\langle S_z \rangle$ and $\langle T_z \rangle$ being the expectation values of spin and magnetic dipole operators for the $3d$ state \cite{PhysRevLett.75.3748}.
Since the $\langle S_z\rangle$ originates solely from the difference in the DOS for the up-spin and down-spin, XMCD from the $S_z$ term is isotropic and thus should cancel out in the coplanar 120$^\circ$ AFM.
On the other hand, since $\langle T_z \rangle$ term gives an angle-dependent XMCD, it becomes possible to separately analyze $\langle S_z \rangle$ and $\langle T_z \rangle$ terms by measuring XMCD spectra at several different angles.

The $T_z$ term in XMCD is defined as the expectation value of the intra-atomic magnetic dipole operator \cite{STOHR1995253}, expressed as 
\begin{equation}\label{t_operator}
t_\alpha=\left[\bm{s}-3(\hat{r}\cdot\bm{s})\hat{r}\right]_\alpha = \sum_\beta Q_{\alpha\beta}s_\beta,
\end{equation}
with $\bm{s}$ and $\hat{r}$ being operators for spin and the unit vector of position, respectively \cite{STOHR1999470}.
Hereafter, the local coordinate ($\alpha,\beta = x,y,z$) is defined based on the orbital and the spin quantization axes.
Equation (3) indicates that $t_\alpha$ is a coupled operator between the spin operator $\bm{s}$ and charge quadrupole operator $Q_{\alpha,\beta}=\delta_{\alpha\beta}-3\hat{r}_\alpha\hat{r}_\beta$ \cite{Oguchi}.
The $t_z$ matrix element calculated on the basis of the spherical harmonic index $m$ and spin index $\sigma$ has spin-diagonal ($s_z$) and spin-flip non-diagonal ($s_\pm= s_x \pm i s_y$) terms \cite{crocombette1996importance}.

The spin-flip term survives only when the spin mixture between two different states ($\Delta m=\pm 1$) is realized in the one-electron eigenstate $\psi$.
Such spin mixing is weak and the spin-diagonal term becomes dominant in the $\langle T_z\rangle$ term for $3d$ transition metal compounds due to the weakness of spin-orbit interaction (SOI). 
Therefore, $\langle T_z\rangle$ mainly reflects the spin-density distribution with the spherical harmonics $Y_2^m$ component.
In a perfect octahedral crystal field $i.e.,$ the $O_h$ site symmetry, $\langle T_z\rangle$ becomes zero as its trace values for both $e_g$ and $t_{2g}$ orbitals are zero.
For example, $\langle t_z\rangle$ values for $e_g$ orbitals are $\langle t_z^{z^2}\rangle = -2/7$ and $\langle t_z^{x^2-y^2}\rangle = 2/7$.
Thus $\langle t_z^{z^2}\rangle+\langle t_z^{x^2-y^2}\rangle=0$ \cite{crocombette1996importance}.  
In other words, a lower-symmetry crystal field is required for the diagonal $T_z$ terms to remain uncancelled \cite{Oguchi}.
For that reason, the $T_z$ term has been utilized as an effective experimental tool in obtaining the magnetic information in quasi-low-dimensional magnets, such as the case of surface magnetism \cite{STOHR1999470,STOHR1995253}.

Accordingly, the $\langle \bm{t}\rangle$ vector depends on the relative angle between the quadrupole quantization axis and the spin.
For example, we consider a hole state $|\psi\rangle$ with the spin canted by the angle $\alpha$ from the $z$ axis on the $3z^2-r^2$ orbital ($Y_2^0$) expressed as
\begin{equation}
|\psi\rangle=
\cos\frac{\alpha}{2}|z^2\uparrow\rangle + \sin \frac{\alpha
}{2}|z^2\downarrow\rangle.
\end{equation}
Assuming that the spin and $\bm{\hat{k}}$ vectors both align within the $xz$ plane, $\langle \bm{t}\rangle = \langle \psi| \bm{t}|\psi \rangle $ can be derived as
\begin{equation}\label{t_vector}
    \langle \bm{t}\rangle = \frac{1}{7}(\sin \alpha, 0, -2\cos\alpha),
\end{equation}
indicating that $\langle \bm{t}\rangle$ is anti-parallel to $\langle \bm{s}\rangle$ at $\alpha = 0$ and $\pi$, parallel at $\alpha=\pm \pi/2$, and perpendicular when $\cos^2\alpha = 1/3$ [see Figs. 1\textbf{c-e}].
Since XMCD can detect the projected component of $\langle \bm{t}\rangle$ to $\bm{\hat{k}}$, represented by
\begin{equation}
\langle t_k\rangle = \frac{1}{7}\sin \alpha\sin \beta-\frac{2}{7}\cos \alpha\cos \beta,
\end{equation}
where $\beta$ is the x-ray incident angle. 
If the spin angle is kept parallel to the x-ray incident direction, $i.e.,$ $\beta=\alpha$, it becomes $\langle t_{k}\rangle = \frac{1}{7}(1-3\cos^2\beta)$.
When $\beta$ is the so-called magic angle, defined as $\cos\beta = \pm \frac{1}{\sqrt{3}}$, $\langle \bm{t} \rangle$ and $\bm{\hat{k}}$ are orthogonal, and the $T_z$ term would disappear in XMCD measurement \cite{PhysRevLett.75.3748}.
It becomes possible to perform an analysis in which only the $S_z$ term is extracted from the XMCD sum rule (Eq. \ref{sum_rule_2}) with removing the $T_z$ term. 

We reveal the $T_z$ term in XMCD for the AMO by considering the magnetic configuration for the Kagom\'{e} AFM.
Here, two typical magnetic structures for the coplanar 120$^\circ$ AFM are discussed.
The unit cell of Kagom\'{e} lattice contains three sublattices (A, B, and C sites) [Fig. 2\textbf{a}]; we set three XY-type spins within the Kagom\'{e} lattice and have 120$^\circ$ mutual spin angle on each sublattice [labeled by colors shown in Fig. 2\textbf{b}].
The Kagom\'{e} lattice has symmetry operations of two translations ($T_1$ and $T_2$), a reflection $\sigma$, and the rotation of $\pi/3$ ($R_6$) about the hexagon center \cite{Messio_PhysRevB83}, as well as a threefold rotational symmetry ($R_3$) about the triangle center.
Two coplanar AFMs, or the so-called $\bm{q}=0$ structures, are shown in Figs. 2\textbf{c} and 2\textbf{d}; 
these two structures are distinguished by vector spin chirality $\bm{\kappa}$ \cite{Hiroi_PRB17}.
In the coplanar spin order, $\bm{\kappa}$ vector for each triangle is normal to the Kagom\'{e} plane, and hence classified by its sign, positive or negative.
The $R_3$ symmetry at the triangle center survives under the magnetic ordering in the case of "positive" vector chirality (PVC) [Fig. 2\textbf{c}], while it breaks in the case of "negative" vector chirality (NVC) [Fig. 2\textbf{d}].
Note that the magnetic structure for Mn$_3$Sn at room temperature corresponds to the NVC \cite{Mn3Sn_Kerr_higo2018large}.

We need to define the quantization axis of the quadrupole moment at each site prior to obtaining $\langle \bm{t}\rangle$ at each magnetic site.
Considering the $R_3$ symmetry at the triangles [Fig. 2\textbf{a}] for magnetic ions at the contact points of two hexagons, we find it preferable to set the local quantization axes $z_A, z_B$, and $z_C$ for respective magnetic ions A, B, and C sites along the directions connecting their centers [Fig. 2\textbf{e}].
This is an important assumption for considering each $\langle\bm{t}\rangle$ in the Kagom\'{e} AFM.
Based on the local coordinates, for example, the $3z^2-r^2$ orbital of each magnetic ion should be oriented as shown in Fig. 2\textbf{f}.
The structure of the $\langle \bm{t}\rangle$ can be estimated in consideration of the orbital arrangement [Fig. 2\textbf{e}] and the magnetic structures [Figs. 2\textbf{c}-\textbf{d}] based on Eq. (\ref{t_vector}).

For the PVC, $\langle \bm{t}\rangle$ also has $R_3$ symmetry as a result of the $R_3$ symmetry for both magnetic moment and electric quadrupole.
Similarly, $\langle \bm{t}\rangle$ would show the 120$^\circ$ antiferroic structure and the $T_z$ term would cancel out regardless of the direction of the local magnetic moment, as depicted in Figs. 3\textbf{a} and 3\textbf{b}.
The NVC is a difference case; $\langle \bm{t}\rangle$ does not have the $R_3$ symmetry due to the broken $R_3$ symmetry $\langle \bm{s}\rangle$, and thus is determined by the relation between $\langle \bm{s}\rangle$ and the quantization axis for each site.
Figures \textbf{3c, 3d} and \textbf{3e} show the NVC AFM when $\theta_B=0^\circ,~45^\circ$ and $90^\circ$, respectively. 
For the NVC, we can see that the $\langle \bm{t}\rangle$ does not cancel but remains to give rise to a ferroic component $\langle \bm{T}\rangle \equiv \sum \langle \bm{t}\rangle$.
As can be seen from Fig. \textbf{3f} showing $\theta_B$ dependence of $\langle \bm{T}\rangle$ components, $\langle {T}\rangle_X$ and $\langle {T}\rangle_Y$, they are antiparallel and parallel to $\langle\bm{s}\rangle_B$ when $\theta_B=0^\circ$ and $\theta_B=90^\circ$, respectively.
The $T_z$ term in XMCD for NVC is proportional to the projection component of $\langle \bm{T}\rangle$ to $\bm{\hat{k}}$ vector.
If the incident direction of x-rays is kept to be parallel to $\langle\bm{s}\rangle_B$, there is a magic angle where  not solely the $S_z$ but the $T_z$ term disappears when $\theta_B$= 45.

The AMO symmetry for those magnetic structures is $T_{X,Y}^\zeta$ for the PVC and $T_{X,Y}^\gamma$ for the NVC \cite{Mn3Sn_MTSuzuki} where the AMO indices depend on the local spin direction.
For the NVC, we can see the index of $T^\gamma_{X,Y}$ AMO being directly linked to the direction of $\langle\bm{t}\rangle$.
Therefore, we can say that the $T^\gamma_{X,Y}$ AMO moment is detectable via the $T_z$ term in XMCD.
This is understandable considering the fact that the AMO can be written by coupled operators of spin and quadrupole moment as Eq. (\ref{t_operator}) \cite{watanabe2017magnetic}.
As mentioned, the $T_z$ term would be zero with the degenerate $e_g$ state.
The site symmetry at Mn ion is $D_{3d}$, in fact which lifts the $e_g$ orbital degeneracy, and thus the $T_z$ term should be allowed in Mn$_3$Sn.
The first-principle calculation indicates that the difference in the DOS between $\pm T_x^\gamma$ AMOs, which is expected to result in a non-zero XMCD signal \cite{Mn3Sn_MTSuzuki}. 
For example, in a cobalt monolayer, the ratio $7\langle T_z\rangle/2\langle S_z\rangle$ is estimated as 0.27, which is not too weak and in fact, cannot be ignored in the sum rule \cite{Oguchi}.
It is also expected to be a sufficiently detectable signal in the Kagom\'{e} NVC AFM as well.
We should note that a quantitative estimation for $T_z$ term is usually difficult because of the strong dependence of $T_z$ on the local geometry of magnetic ions.

We have revealed that XMCD derived from the pure $T_z$ term is observable for the NVC structure, and related to the $T_{X,Y}^\gamma$ AMO in the Kagom\'{e} AFM.
Use of this will be an effective approach for ferroic ordering of AMO in AFMs with the breaking of TRS.
If the $\langle \bm{t}\rangle$ vector shows antiferroic order, it can be detected by resonant x-ray scattering.
Although the SOI effect is ignored in the present study, extracting information from the non-diagonal $T_z$ term might be also possible.
For example, a higher multipole moment caused by a mixture of $e_g$ and $t_{2g}$ orbitals, such as a magnetic triakontadipole ($2^5$-pole) moment \cite{ishihra_EI}, is also expected to be observable via the XMCD $T_z$ term.

\bibliography{bib2}

\begin{figure}[t]
\includegraphics[width=10.cm]{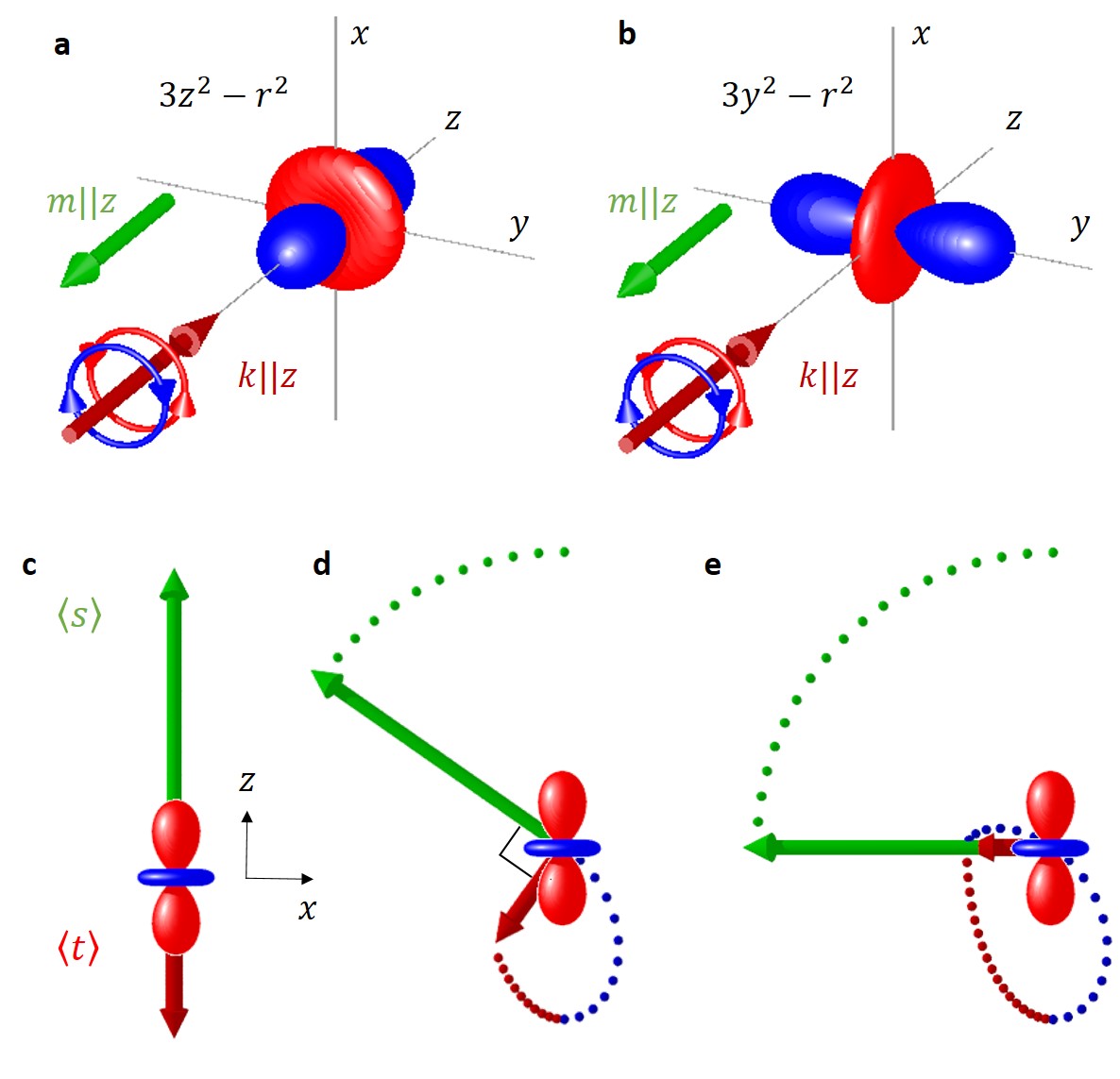}
\caption{\label{fig:epsart} \textbf{Anisotropic XMCD and $\langle \bm{t} \rangle$ vector for $3z^2-r^2$ orbital:} Even under the same condition of $\bm{m}||\bm{k}||z$, the signal intensity of XMCD differs between orbitals \textbf{a,} $3z^2-r^2$ and \textbf{b,} $3y^2-r^2$ due to orbital anisotropy within the $xy$ plane. 
\textbf{c-e,} Expectation value of $\langle \bm{s}\rangle$ and $\langle \bm{t}\rangle$ vectors at three representative spin angles $\alpha$ for the $3z^2-r^2$ orbital. 
Green (Red) arrows and dots indicate the expectation vector and its trajectory for $\langle \bm{s}\rangle$ ($\langle \bm{t}\rangle$). 
Blue dots are the trajectory of the projected vector of $\langle \bm{t}\rangle$ to $\langle \bm{s}\rangle$ directions.
For a single spin rotation, $\langle \bm{t}\rangle$ draws an elliptical orbit as it rotates in the reverse direction.
When the x-ray incidence is kept parallel to $\langle \bm{s}\rangle$, the $T_z$ term in XMCD is zero at the so-called magic angle, as shown in \textbf{d}, where $\langle \bm{t}\rangle$ and $\langle \bm{s}\rangle$ become orthogonal to each other.}
\end{figure}

\begin{figure*}[t]
\includegraphics[width=15cm]{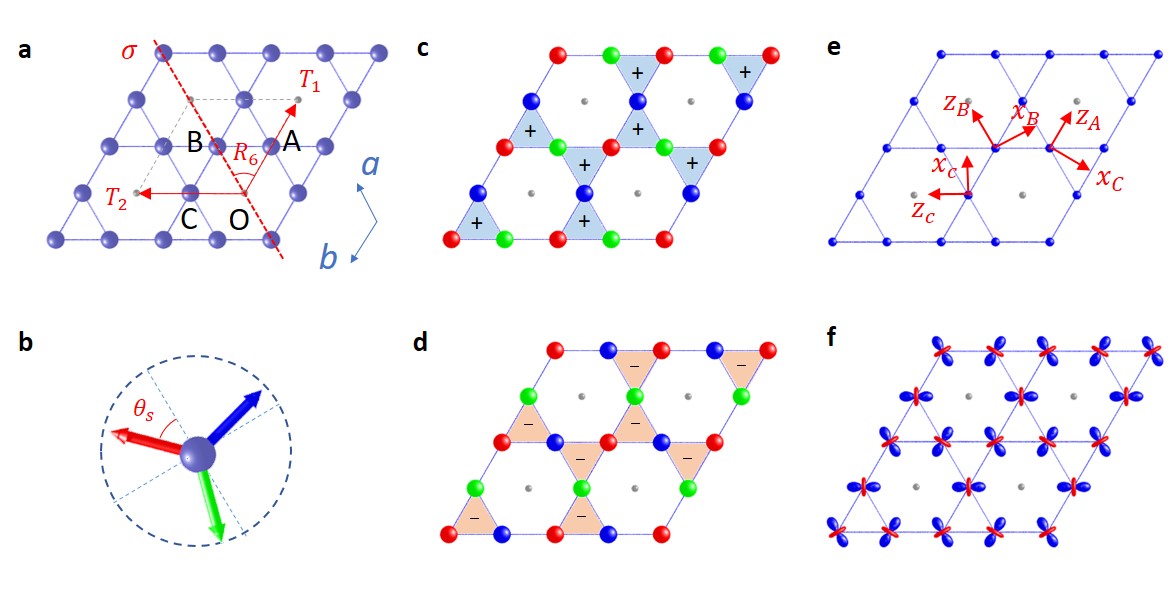}
\caption{\textbf{Lattice structure and magnetic orders on the Kagom\'{e} lattice:} \textbf{a}, Three sublattices (labeled as A, B, and C), and lattice symmetries of two translations ($T_1$ and $T_2$), reflection $\sigma$, and rotation of $\pi/3$ ($R_6$).  \textbf{c} - \textbf{f}, Coplanar 120$^\circ$ magnetic orders labeled by colors as depicted on \textbf{b}. \textbf{c} and \textbf{d} are $\bm{q}=0$ structures with positive and negative vector spin chiralities (PVC and NVC), respectively. The signs ($"+"$ and $"-"$) indicate directions of the vector spin chirality for each triangle \cite{Hiroi_PRB17}. \textbf{e}, Quantization axis for quadrupole moment and \textbf{f}, the $3z^2-r^2$ orbital configuration based on the local coordinate at each magnetic site.}
\end{figure*}

\begin{figure}[t]
\includegraphics[width=16cm]{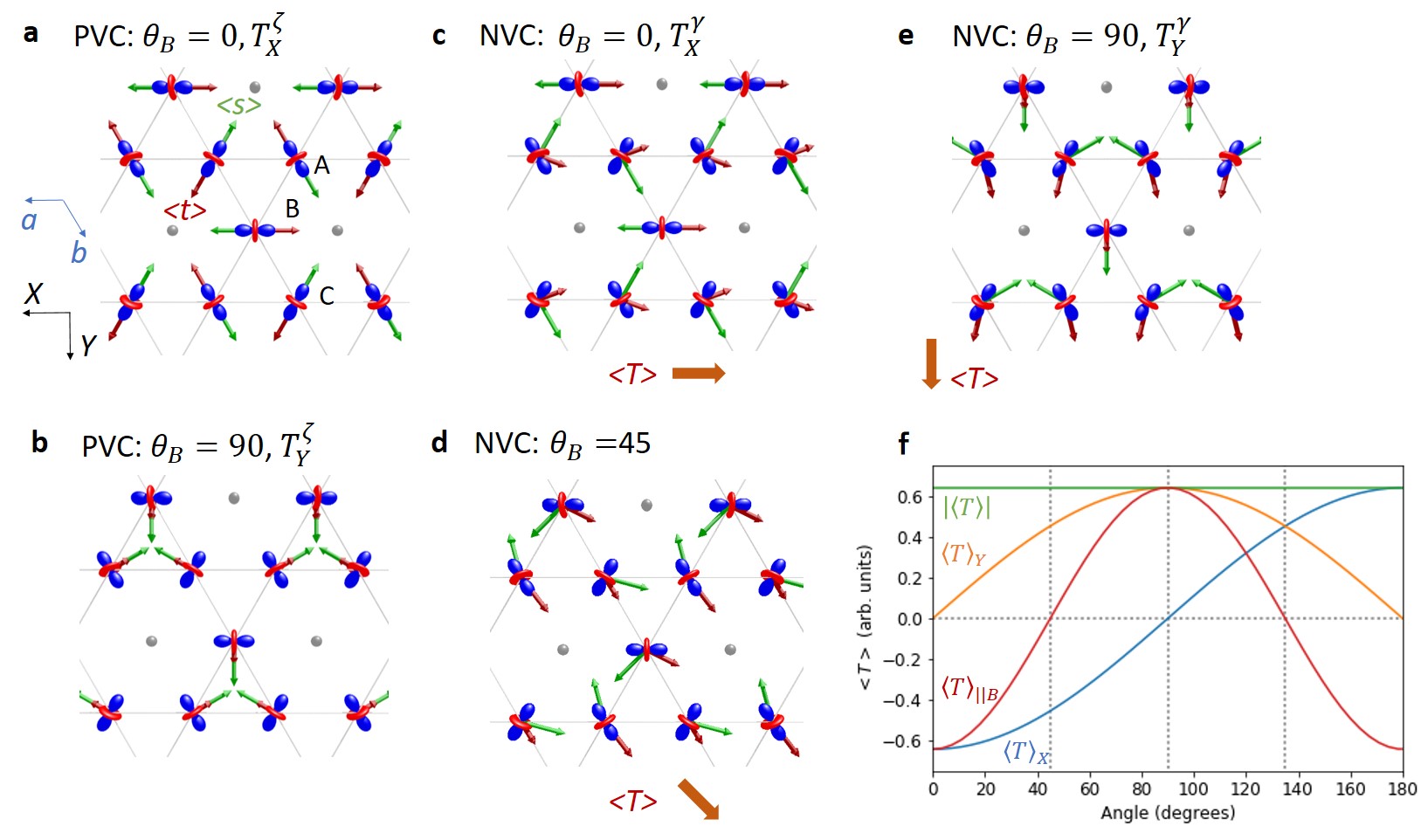}
\caption{\label{fig:epsart} \textbf{$\langle \bm{T}\rangle$ vector in coplanar 120$^\circ$ AFM:} Triangle antiferromagnetic order for \textbf{a}, $T_X^\zeta,~\theta_B = 0^\circ$ and \textbf{b}, $T_Y^\zeta,~\theta_B = 90^\circ$ symmetry of the AMO in the $q=0$ PVC, and for \textbf{c}, $T_X^\gamma,~\theta_B = 0^\circ$, \textbf{d}, $\theta_B = 45^\circ$, and \textbf{e}, $T_Y^\gamma,~\theta_B = 90^\circ$ in the NVC.
The $X$ and $Y$ axes correspond to $[2\bar{1}\bar{1}0]$ and $[01\bar{1}0]$ axes in Mn$_3$Sn, respectively. 
$\theta_B$ indicates the spin angle at B-site.
Both $\langle \bm{s}\rangle$ and $\langle \bm{t}\rangle$ cancel out in the PVC structure. In contrast, uncompensated $\langle \bm{T}\rangle = \sum \langle \bm{t}\rangle$ emerges in the NVC even without net magnetization. \textbf{f}, The spin angle $\theta_B$ dependence of $\langle T \rangle$ components; $\langle T\rangle_X$, $\langle T\rangle_Y$, $|\langle T\rangle|$, and $\langle T\rangle_{||B}$ are $X$ and $Y$ components, magnitude, and projection on the B-site spin direction of $\langle \bm{T}\rangle$ vector, respectively.}
\end{figure}

\clearpage
\section*{Author contributions}
Y.Y. conceived of the presented idea, and developed the theory and performed the analytic calculations. H.N. and T.A. verified the analytical methods. All authors discussed the results and contributed to the final manuscript.

\section*{Acknowledgement}
This work was supported in part by Materials Research by the Information Integration Initiative (MI2I) project of the Support Program for Starting Up Innovation Hub from the Japan Science and Technology Agency (JST), by PRESTO Grant Number JPMJPR177A, and by Grant-in-Aid for Scientific Research Nos. JP16H05990, JP15H05885(J-Physics), JP19H01835, and JP19H05826 from the Japan Society for the Promotion of Science (JSPS).

\end{document}